\documentclass[apj]{emulateapj}
\journalinfo{\textsc{The Astrophysical Journal}}
\slugcomment{Received 2014 May 29; Accepted 2014 October 8}
\usepackage{url}

\usepackage{apjfonts}
\usepackage{natbib}
\usepackage{booktabs}

\usepackage[colorlinks,
        hypertexnames=false,
        pagebackref=true,
        linkcolor=red,    
    citecolor=blue,        
    filecolor=magenta,     
    urlcolor=red          
]{hyperref}                
\usepackage{amsmath}
\usepackage{tipa}

\newcommand{\Ha}{H$\alpha$}

\shorttitle{Flux Rope Initiation and Its Eruption to CME}
\shortauthors{Vemareddy and Zhang}

\begin{document}
\title{Initiation and Eruption Process of Magnetic Flux Rope from Solar Active Region NOAA 11719 to Earth Directed-CME}
\author{P.~ Vemareddy$^1$, and J.~Zhang$^2$}
\affil{$^1$Udaipur Solar Observatory, Physical Research Laboratory, Badi Road, Dewali, Udaipur-313 001, India.}
\affil{$^2$ School of Physics, Astronomy and Computational Sciences, George Mason University, Fairfax, VA 22030, USA}
\email{vema@prl.res.in}


\begin{abstract}
An eruption event launched from solar active region (AR) NOAA 11719 is investigated based on coronal EUV observations and photospheric magnetic field measurements obtained from Solar Dynamic Observatory. The AR consists of a filament channel originating from major sunspot and its south section is associated with inverse-S sigmoidal system as observed in AIA passbands. We regard the sigmoid as the main body of the flux rope (FR). There also exists a twisted flux bundle crossing over this FR. This overlying flux bundle transforms in shape similar to kink-rise evolution which has correspondence with rise motion of the FR. The emission measure and temperature along the FR exhibits increasing trend with its rising motion, indicating reconnection in the thinning current sheet underneath the FR. Net magnetic flux of the AR evaluated at north and south polarities showed decreasing behavior whereas the net current in these fluxes exhibits increasing trend. As the negative (positive) flux is having dominant positive (negative) current, the chirality of AR flux system is likely negative (left handed) to be consistent with the chirality of inverse S-sigmoidal FR. This analysis of magnetic fields of source AR suggest that the cancelling fluxes are prime factors to the monotonous twisting of the FR system reaching to a critical state to trigger kink instability and the rise motion. This rise motion possibly led to onset of torus instability resulting in Earth-directed CME and the progressive reconnection in thinning current sheet underneath the rising FR leads to M6.5 flare.
\end{abstract}


\section{Introduction}
\label{Intro}

It is well accepted that solar eruptions are magnetically driven events influencing the physical processes and phenomena in the interplanetary medium in a wide range and the space weather near the Earth. During the eruptions, a commonly observed physical structure is magnetic flux rope \citep{burlaga1988,lepping1990,zhangj2012}. It is modeled as a volumetric current channel with helical field lines wrapping around its central axis. Filaments/prominences seen in \Ha~\citep{Vemareddy2011}, and coronal features of sigmoids are often considered as flux ropes in ARs. The preexistence, formation and role of magnetic flux rope are not well understood. It has been traditionally believed that magnetic reconnection plays the main role and flux rope is treaded as secondary, but different views exist.

Recent direct evidences of the flux rope come into existence as a conspicuous hot channel structure in the inner corona of the AR before and during a solar eruption \citep{Vemareddy2011,zhangj2012}. This channel initially appears as a twisted and writhed sigmoidal structure in high temperature passbands. The channel evolves toward a semi-circular shape in the slow rise phase and then erupts upward rapidly in the impulsive acceleration phase, producing the front-cavity-core components of the resulting CME \citep{zhangj2012}. The role of this hot channel in the eruption process is similar to that of flux rope in many theoretical and numerical models \citep[e.g.,][]{chenj1996,chenpf2000,linj2000,torok2005,kliem2006,aulanier2010,fany2007,fany2010,olmedo2010} reproducing key features of eruptions.

As most sigmoid eruptions lead to CMEs \citep{canfield1999}, they are interpreted by two different view points of the same structure as sheared arcade or flux rope. The sheared arcade model assumes that the sigmoid is composed of sheared and twisted core field of the AR, and the internal runaway tether-cutting reconnection is responsible for the eruption \citep{moore2001,liur2010}. Whereas, the flux rope kind of models argue that the evolution of a sigmoidal AR with flare and CME is due to twisted magnetic flux rope that emerges and equilibrates the overlying magnetic field structure \citep{gibson2002,gibson2006}.

Many theoretical and numerical models were formulated to predict the triggering mechanism of eruptions that involve filaments, prominences or sigmoids. Of them  the flux rope based models seem successful in reproducing eruption features. Flux rope would lose equilibrium after reaching a critical height, forming current sheet beneath as proposed by \citet{forbes1991, priest2002}. Kink instability can initiate the the rise motion when twist of flux rope exceeds a threshold \citep{torok2004}. Torus instability \citep{kliem2006} is believed to be the mechanism of ultimately triggering the eruption when there is rapid decline of the background field in the direction of the expansion of the flux rope. For the evidence of the kink instability in coronal data, \citet{rust1996} compared the shape of 49 transient bright sigmoid structures to the geometry of a helically kinked flux rope. They assumed this rope to have 2$\pi$ of writhe over its length and showed that the distribution of sigmoids as a function of aspect ratio falls off abruptly below the threshold for the m=1 kink mode in this model. From this observation they inferred that kinking is the cause of eruption. Similarly, the pitch angle of erupting prominences with helical structure \citep{vrsnak1991} was found to be greater than 2.5$\pi$ and no prominence eruption observed with pitch angle less than 2$\pi$. However, signatures of required amount of twist for the kink-instability prior to the eruption remain difficult to determine \citep{leamon2003, leka2005}.

In the present paper, we studied an eruption event involving sigmodal flux rope using high cadence, high resolution observations from Solar Dynamic Observatory in view of present ambiguities in the physics of flux rope formation and eruption as discussed above. Our main focus in this work is on the questions: Whether the flux rope pre-existed or it formed during eruption? At what stage of the eruption the flux rope formed? What happens to the flux rope during the eruption? What evolving conditions of magnetic fields in the host active region led to the flux rope eruption? Answers to these key questions help address the formation, initiation mechanisms of flux rope eruptions. The observational analysis of such events provides better insight on the formation and destabilization of flux ropes in the host active regions and supplements the key information to constrain and reformulate the existing models of eruptions that rely on flux ropes. The event studied here is unique from the usually observed sigmoidal structures formed in filament channels that fully involved in flux rope formation at the time of onset of the eruption. The sigmoid associated with a section of the filament channel destabilized leading to eruption and suggests for its dependence on local factors of magnetic fields.

Morphological investigation of the FR evolution to identify its preexistence/formation, initiation mechanism, is presented in Section~\ref{MorphProp}. Details on the kinematics of the FR are given in Section~\ref{Kinets} and its thermal properties like temperature and density structures are presented in Section~\ref{TherProp}. A possible triggering mechanism of the eruption based on photospheric magnetic field observations is addressed in Section~\ref{MagProp}. We conclude the manuscript with summary and discussion in Section~\ref{Summ}.

\begin{figure*}[!htp]
\centering
\includegraphics[width=.92\textwidth]{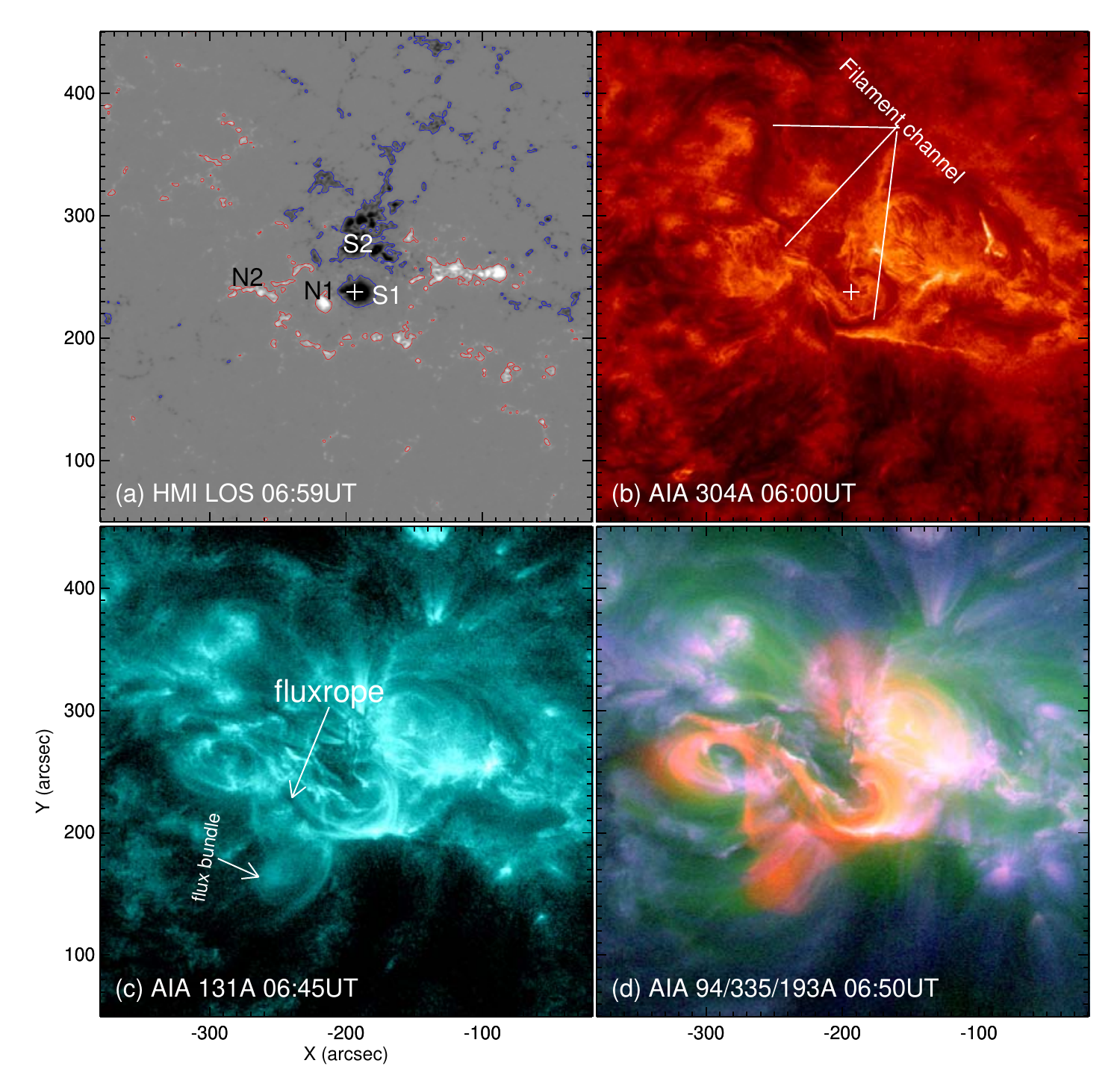}
\caption{The eruptive active region NOAA 11719 and sigmodal flux system. a) Magnetic flux distribution in HMI line-of-sight magnetogram. Contours at $\pm$150 G in red/blue are shown. Major flux regions are marked as N*/S* to refer north/south polarities. b) Observations in AIA 304~\AA, showing filament channel. It originates from sunspot S1 and extends to north part of diffused positive flux of the AR. Note the helical like structure at the middle of filament channel. c) Corona in hot passband of 131~\AA. An inverse-S shaped sigmoidal FR is observed to associate with the south section of the filament channel. It connects S1 flux and N2 flux regions and had over-crossing loop as flux bundle, which is hardly visible d) the proximity of FR system in composite image constructed from AIA observations in 94 (red), 335 (blue), 193~\AA~(green) wavelength passbands. Note the clear sigmoidal structure. And the overlying loops also as hot as main sigmoid.}
\label{Fig1_Mos1}
\end{figure*}

\begin{figure*}[!htp]
\centering
\includegraphics[width=1.0\textwidth]{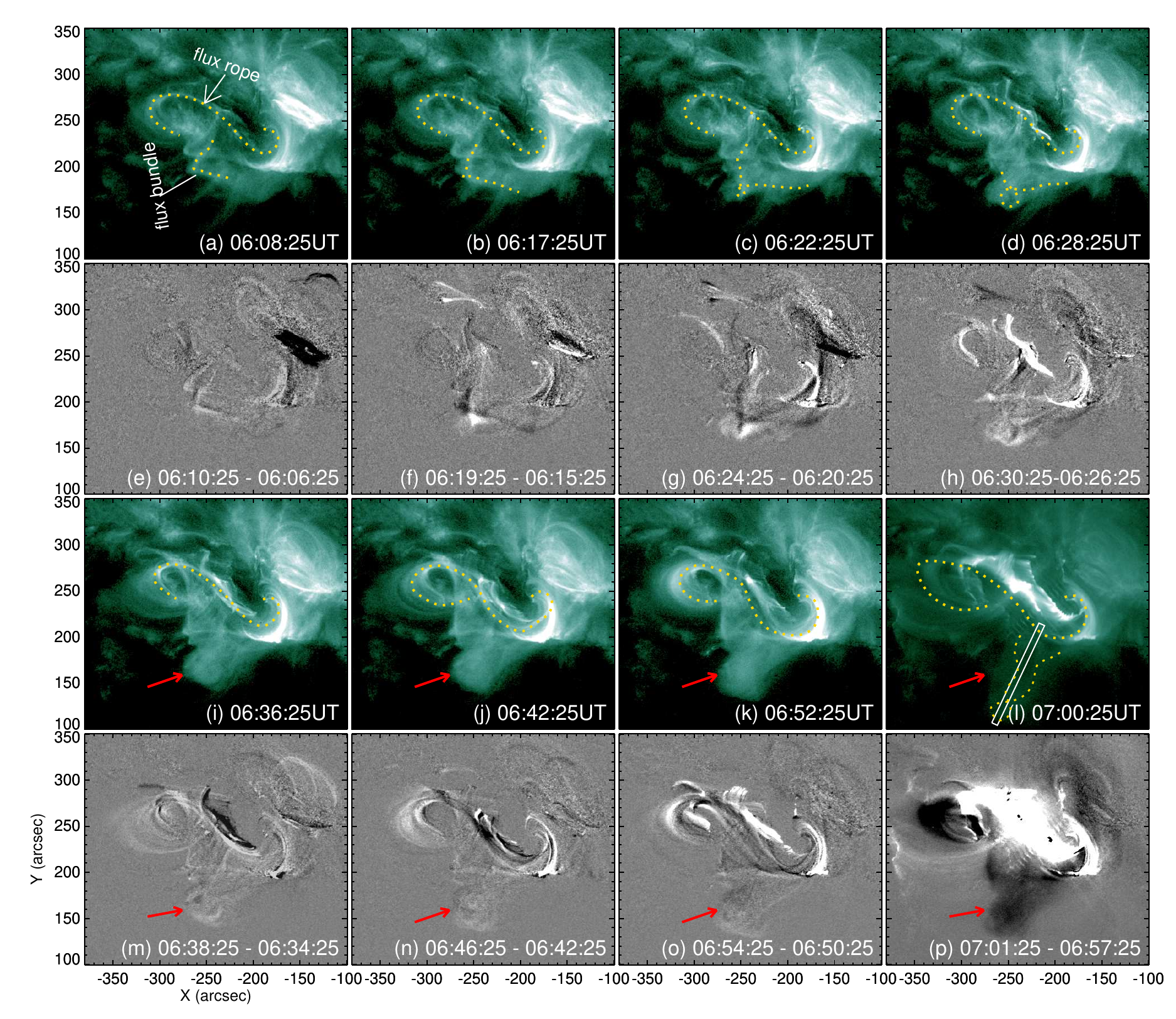}
\caption{The pre-eruption scenario of sigmoidal AR 11719 in AIA 94\,\AA~observations.  (a-d) Morphological evolution of sigmoidal FR system depicting the initiation of rise motion.  The trace of flux rope is identified and indicated by dotted yellow curve.  The over crossing flux bundle shows interesting evolution pattern. We use difference images as plotted in (e-h) to trace the flux bundle (highlighted by yellow dotted curve). They indicate the transformation of the flux bundle from V- to \textgamma- shape structure. (i-l) Snapshots in further evolution period delineate upward rise motion of the flux rope. The overlying flux bundle also rises in successive stages but its apex (marked by red arrow) is smeared due to diffused plasma filling appearing as plasma blob of closed loop structure. Even difference images preclude to reveal the details of the loops within the apex structure of the flux bundle. The rise motion of the FR led to commencement of a flare from 06:55 UT and its eventual eruption by 07:00 UT. A rectangular slit is shown for a study of kinematic evolution of this flux bundle (panel l). Flare ribbons formed beneath the well lifted FR as observed in the last panel. (Please also refer to the accompanied animation)}
\label{Fig1_Mos2}
\end{figure*}

\section{Observational Data}
\label{ObsData}
The active region in which the eruption occurred on April 11, 2013 was NOAA 11719, located at N07$^{o}$E13$^{o}$ on the solar disk. This event was extensively observed by the instruments, \textit{Atmospheric Imaging Assembly} (AIA; \citealt{lemen2012}) in UV and EUV wavelengths providing coronal imaging information and \textit{Helioseismic Magnetic Imager} (HMI; \citealt{schou2012}) in magnetically sensitive Fe {\sc i} line at 6173\,\AA~providing photospheric magnetic field measurements, on board Solar Dynamic Observatory. Using standard procedures in SolarSoft, the images of corona are processed and aligned to magnetograms which are already transformed to solar disk center \citep{calabretta2002} after azimuthal ambiguity resolution \citep{leka2009} of horizontal field vectors derived from Very Fast Inversion of the Stokes Vector  \citep[VFISV;][]{borrero2011} inversion algorithm. The analysis results of the FR eruption from these observations are described in the following subsections.
\section{Data Analysis and Results}
\subsection{Morphological Evolution}
\label{MorphProp}
Figure~\ref{Fig1_Mos1} illustrates the pre-eruptive situation of the AR in different wavelengths from photosphere to extended corona.

The magnetic flux distribution of the AR from HMI is plotted in Figure~\ref{Fig1_Mos1}(a). It shows major sunspot of negative flux (S1) surrounded by parasitic, diffused positive flux. The sunspot is flanked by small sunspot of positive polarity and sunspot group of negative polarity patch. This distribution of magnetic flux concentrations corresponds to multi-polar configuration and stands in the complexity of $\beta\gamma$ classification. From AIA 304\,\AA~passband that gives plasma information from chromosphere and transition region He {\sc ii} ($\sim$10$^4$K), it clearly shows existence of dark filament channel approximately oriented in north-south direction (panel~\ref{Fig1_Mos1}(b)). One of this filament legs anchors from sunspot of negative polarity flux S1, and extends to northern plague regions of dispersed positive fluxes. Filament channels containing cool plasma material compared to surrounding temperature will appear as dark trace, thus visible in 304\,\AA~or \Ha~as reported in \citet{Vemareddy2011}. This filament channel swirls around the sunspot in clockwise fashion, which gives its appearance as inverse-S geometrical shape to it. The middle part of this channel is helical where north and south sections connected which is unclear from static images. As most filaments do, this filament channel also traces approximately amid opposite polarity fluxes.

However, in hotter passbands of AIA 131 ($\sim$10\,MK), 94\,\AA~($\sim$6\,MK), this north-south-oriented filament channel is invisible because of its low temperature conditions. Instead, these images show a structure with inverse S-shaped coronal loops filled with bright plasma in a different location. This sigmoidal flux system oriented roughly along East-West line, connecting S1 and N2. It spatially merges with the south section of the filament channel from S1, but deviates from middle section of the filament channel. Because of this flux system visibility in high temperature passbands, we infer that the flux system is situated higher up in the corona above the filament channel. These observations are generally supportive with the view to explain the magnetic topology of filament systems associated with sigmoidal configuration that the filament is supported by twisted flux rope against gravity \citep{rust1994, gibson2002}. In this picture, the filament material is held in the dips of twisted field lines of the flux rope along the magnetic inversion line. In this present scenario of observations, we infer that the observed filament consists of two separate sections supported by two different twisted flux systems. The north section of filament channel remains to be stable without noticeable geometrical or temporal changes from the past two days and completely unrelated to this eruption event. The sigmoid associated with the south section of the filament channel, is involved in this eruption. The visible features of this inverse-S sigmod and the underlying filament will be focused in this study.

The active region magnetic flux system associated with the sigmoid and the south section of the filament is having a twisted flux sub-system in the form of hot plasma loops (panel~\ref{Fig1_Mos1}(c)) crossing over the middle section of the sigmoid. We refer this over-crossing flux sub-system as a twisted flux bundle throughout this paper, in order to distinguish from the underlying sigmoid (flux rope) and the filament. While also made of twisted flux bundles, sigmoids and filaments are well defined in observations and physical terms. The photospheric connectivity of the twisted flux bundle in the above is not clear because of faint emission but they seem to be connected to sunspot S1 flux and positive diffused flux of N2. We regarded this whole system observed in hot passbands as a complex twisted flux system including multiple sub-systems: the sigmoid as the main body of the flux rope (FR) and the kinked hot loops crossing over this FR as a distinct twisted flux bundle. The sequential phases of evolution of this multiple-component system prior to eruption are discussed as follows.

From the animations of 131\,\AA~images, the sigmoid starts appearing from 06:25\,UT onwards surrounded by the twisted flux bundle in the form of plasma loops. With time, the east lobe of the sigmoid starts rising with increased spread of brightness along the central core of the sigmoid. As a result, the central core becomes less bright giving a coherent FR appearance to it. This can be deliberately noticed after 06:45\,UT as depicted in Figure~\ref{Fig1_Mos1}(c). As the sigmoid progressively reached to higher heights, there observed enhancing bright emission that prevails throughout the whole flux system as noticed in hotter 94\,\AA, 131\,\AA~passbands. This phase immediately follows triggering of the fast flare magnetic reconnection, thus entering into a phase of flare of magnitude M6.5 from 06:55\,UT (flare onset) onwards. The flare is associated with bright ribbons on either side of the channel. These observations are similar to \citet{zhangj2012} reporting FR visibility in hot channels prior to the onset of the flare.

The observability of sigmoid in 94\,\AA~prior to eruption indicates heating to at least EUV range, if not soft X-ray. During the dynamical evolution, the FR is found to associated with sigmoidal current sheet as seen in many numerical MHD simulations \citep{gibson2004, gibson2006}. The persistent reconnection in this current sheet and the emission of resultant heated plasma during rising motion of FR is suggested to be responsible for the brightening of observed transient sigmoidal structure.

Composite images of corona, transition region made from the images of AIA 335\,\AA, 94\,\AA, 193\,\AA~ passbands provide abundant signatures of the interconnection of the two sub-flux systems (panel~\ref{Fig1_Mos1}d). The twisted flux bundle in the above also indicates the high temperature conditions. \citet{titov1999} already suggested that the transient sigmoidal brightening outlines the magnetic structure with enhanced current density. The magnetic structure in the sigmoidal FR system comprises of helical field lines winding about the central axis although we cannot observe these structures in our images. The cooling of hot plasma residing at different flux shells of this FR system will be at different temperatures due to thermal insulation by magnetic field and outlines the magnetic structure when observed in different wavelengths. Therefore, the plasma in the sigmoidal FR is likely multi-thermal and is very helpful to reveal magnetic topology of the FR \citep{tripathi2009}. The composite snapshots (panel~\ref{Fig1_Mos1}(d)) well depict this picture of sigmoid structure which is generally observed in soft x-rays \citep{rust1996} as a FR system with central magnetic axis threaded by helical field lines.

In 94\,\AA~images (Figure~\ref{Fig1_Mos2}(a-d)), the same picture of evolution can be framed from the diffused hot plasma in the FR structure. A faint twisted hot loop that is likely anchored from the sunspot overly the FR at the middle. It gives the impression that the FR is tied by it to keep the system stable. Their different geometrical appearance suggests that FR and the flux bundle possess different magnitudes of twists albeit in the same sense of sign. They are likely having left-handed chirality field, in order to give inverse-S orientation in projection, following hemispheric chirality rule as observed for filaments or sigmoids \citep{rust1994, martin1994, gibson2006}.

From animations, evolution of the flux bundle in 94\,\AA~observations exhibits interesting morphological changes associated with geometrical transformation. Supplement movies are provided for a better viewing of the dynamic process. Static images hardly reflect rise motion of flux rope and the flux bundle. To illustrate their dynamics, we use running difference images of AIA 94 observations (panels~\ref{Fig1_Mos2}(e-h), (m-p)), which enable us to follow their trace by a device-cursor procedure. We highlight the trace of both FR and the flux bundle by yellow dotted line in different frames. The flux bundle appeared as $V$ shaped structure in the early phase (panel~\ref{Fig1_Mos2}(a)). At a later stage, it turned to cusp-like structure ($\gamma$), which then evolved to \textgamma-shaped structure by 06:30 UT. Within this 25 minute duration of evolution, the apex of the flux bundle made a rotation to form a closed loop structure above its legs. Thereafter, the apex is spread with hot diffused plasma because of which it appears as a plasma blob. Although there is certain dynamical activity of flux bundle within this blob, it is not clear whether closed loop structure further rotated. At the time of the flare at 07:00\,UT, this flux bundle structure is stretched to extended heights (see its approximate trace in panel~\ref{Fig1_Mos2}(h)). 

The kind of evolution transforming the flux bundle from  $V$- through $\gamma$- to \textgamma-shape is presumed to be a signature of writhing of the flux bundle. When the twist in the flux rope exceedingly increasing, the axis of the rope loops around itself  resulting in writhing and the flux rope experiences kinking \citep{berger1999}. When observed in a particular orientation, either $\gamma$- or \textgamma- shaped flux ropes will be realized due to projection. These observational signatures of kink-rise evolution are usually seen in the H$\alpha$ observations of filaments at the limb side (e.g., \citealt{alexander2006, liur2007}). In concurrent with theory, well-established observational definitions of kinking of filaments (flux ropes) are furnished in \citet{gilbert2007}. In our case, the flux bundle is oriented as if we see it on the limb facilitating us to visualize its kinked behavior. It evolved to a clear kinked structure with closed loop above the legs. Although there remains a difficulty with diffused nature of the flux bundle in addition to prevailed projection effects which may result in misleading interpretation of the observations, the inferred evolution of the flux bundle is similar to writhing motion. The availability of magnetic twist required for such coronal observational signatures of flux rope structures can be estimated from the magnetic field observations at the photosphere, which in turn confirms this morphologically observed geometrical evolution of this flux bundle as a signature of magnetic twist. In the case of limb events, as magnetic field observations are not suitable, simultaneous line-of-sight velocity data will be helpful to confirm the writhing motion of the flux rope which is a signature of magnetic twist (e.g., \citealt{alexander2006, liur2007}). 

The rise motion of flux bundle corresponds well with that of the underlying flux rope. During the writhing phase of the flux bundle (06:06-6:30 UT), the flux rope is observed with slow rise motion. It then follows further upward motion of both the flux rope and the flux bundle during the time 06:30-06:55 UT, which is followed by their drastic rise phase subsequently the eruption. Therefore, the kink-rise evolution of the flux bundle may be the signature of availability of critical twist (see Section~\ref{MagProp}) and its co-temporal rise motion with FR implies either the rising flux bundle allowed the underlying FR to rise and/or the FR itself had exceeding twist so that it itself can rise. Both these assumptions need to be supported by the observations of the sufficient twist in the AR flux system by means of photospheric magnetic fields. Tentatively, we believe that the rise motion of the main flux rope caused by the kink instability of overlying flux bundle eventually led to the onset of the torus instability at the time of the flare onset, resulting in a CME. The CME event\footnote{\url{http://cdaw.gsfc.nasa.gov/CME_list/UNIVERSAL/2013_04/univ2013_04.html}} is a halo one heading towards Earth at a linear speed of 861 km/s.

After this initiation phase, the whole system enters into an impulsive phase, the main phase of the eruption, characterized by the fast acceleration of the FR and the main release of the flare energy. This impulsive phase is then followed by the post flare phase with the appearance of potential-like coronal arcade loops. The post flare phase extends for about four hours having peak EUV emission at about 07:45\,UT, and the north section of the filament remains unaffected.

\begin{figure*}[!htp]
\centering
\includegraphics[width=.72\textwidth]{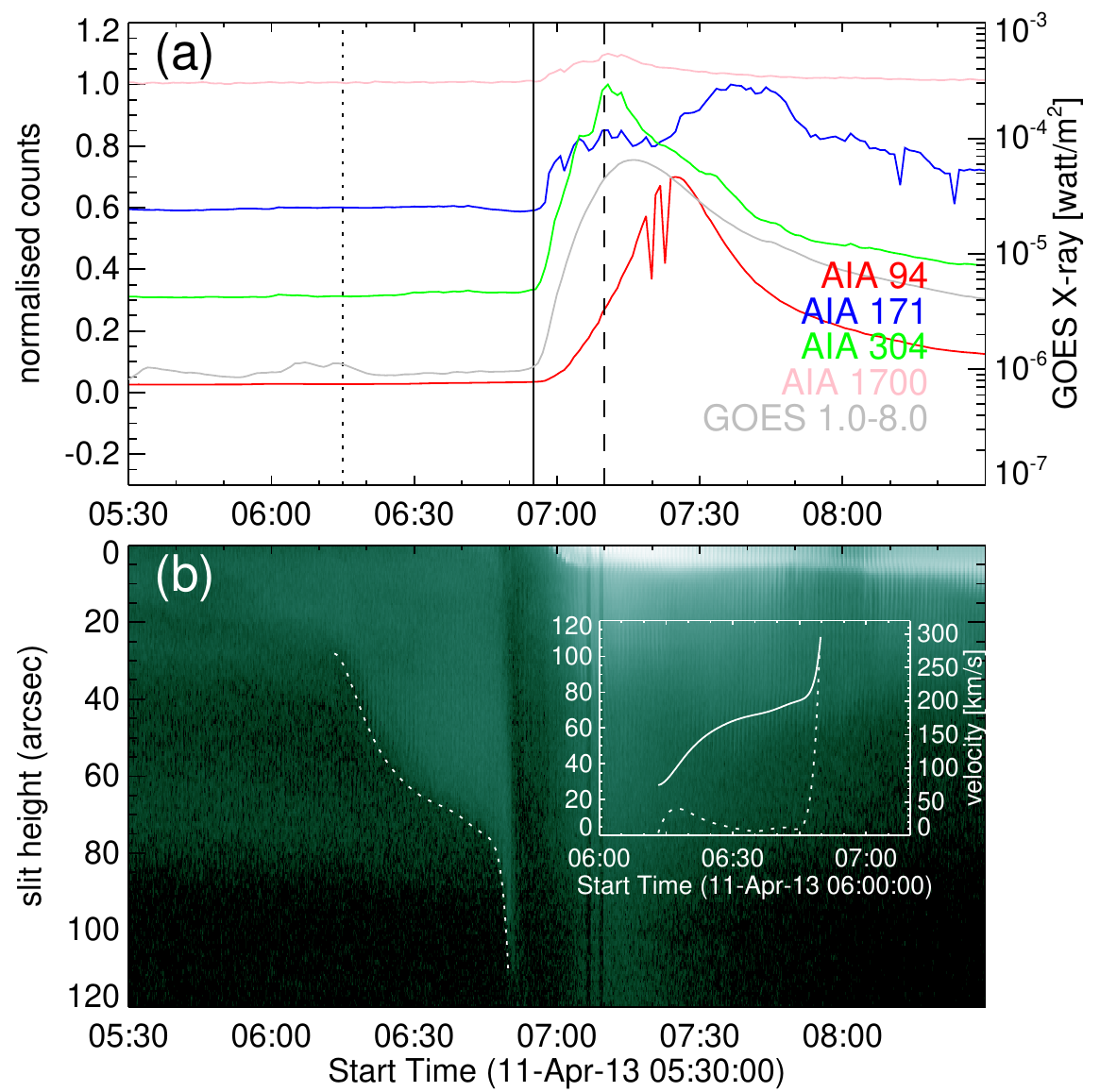}
\caption{ a) Light-curves of the AR in different wavelengths. Dotted vertical line refers to start time of flux bundle rise as seen in AIA 131, 94\,\AA~channels. The solid and dashed vertical lines indicate the start and peak time of the flare that is preceded by the FR eruption, respectively. b) Slice-time image constructed from the slit shown in Figure~\ref{Fig1_Mos2}(l), to provide information about rise-motion of flux bundle as a proxy for the kinematics of the main FR. Onset of the flare is concurrent with fast acceleration of the FR. In the inset, we plot height, velocity of the flux bundle with time.}\label{Fig_LigCur_Sl}
\end{figure*}

\subsection{Kinematic Analysis}
\label{Kinets}
In order to know details of the initiation mechanism, we carried out kinematic analysis of the twisted flux bundle, which also represents as proxy for the main FR rise motion. A vertical slit across the apex of the flux bundle (Figure~\ref{Fig1_Mos2}(l)) is remapped in each time frame of 94\,\AA~images to construct slice-time map as plotted in Figure~\ref{Fig_LigCur_Sl}(b). Because of inclined orientation, there may be projection effects crept in the estimation of height of the flux bundle, so the values indicate only approximations. While rising, the flux bundle is moving down towards south, giving a distinct intensity pattern as indicated by dotted curve in time-slice image. The curved path is fitted with a polynomial function (of degree 13 containing only odd terms and gives a goodness of fit as 0.88) and derived velocity profile as depicted in the inset plot.

Within the heights of low corona that is captured in hot channels, the flux bundle rising motion can be divided into three distinct phases as analyzed from the plot:  a moderate rise of flux bundle at a velocity of 25\,km\,s$^{-1}$ in its writhing phase 06:15-06:30 UT, a slow rise motion at a velocity of 10~km\,s$^{-1}$ during 06:30-06:45 UT, and fast acceleration during its stretching phase from 06:45 onwards with velocity reaching to 270 km\,s$^{-1}$. The last phase is followed by eventual eruption, and a flare of electromagnetic radiation. The light curves of the flare in prominent wavelengths are plotted in Figure~\ref{Fig_LigCur_Sl}(a). The time profile of flux in 304, 1700\,\AA~follows the soft x-ray flux with co-temporal peak flux at 07:10\,UT. Similarly the emission in 171\,\AA~commensurately increases with soft x-ray flux and in addition a late peak emission at about 07:35\,UT occurred corresponding to post flare loops. The peak emission from 94\,\AA~filter delayed by 15 minutes to that of soft x-ray flux but earlier by 10 minutes to the second peak of 171\,\AA.

\begin{figure*}[!ht]
\centering
\includegraphics[width=.97\textwidth]{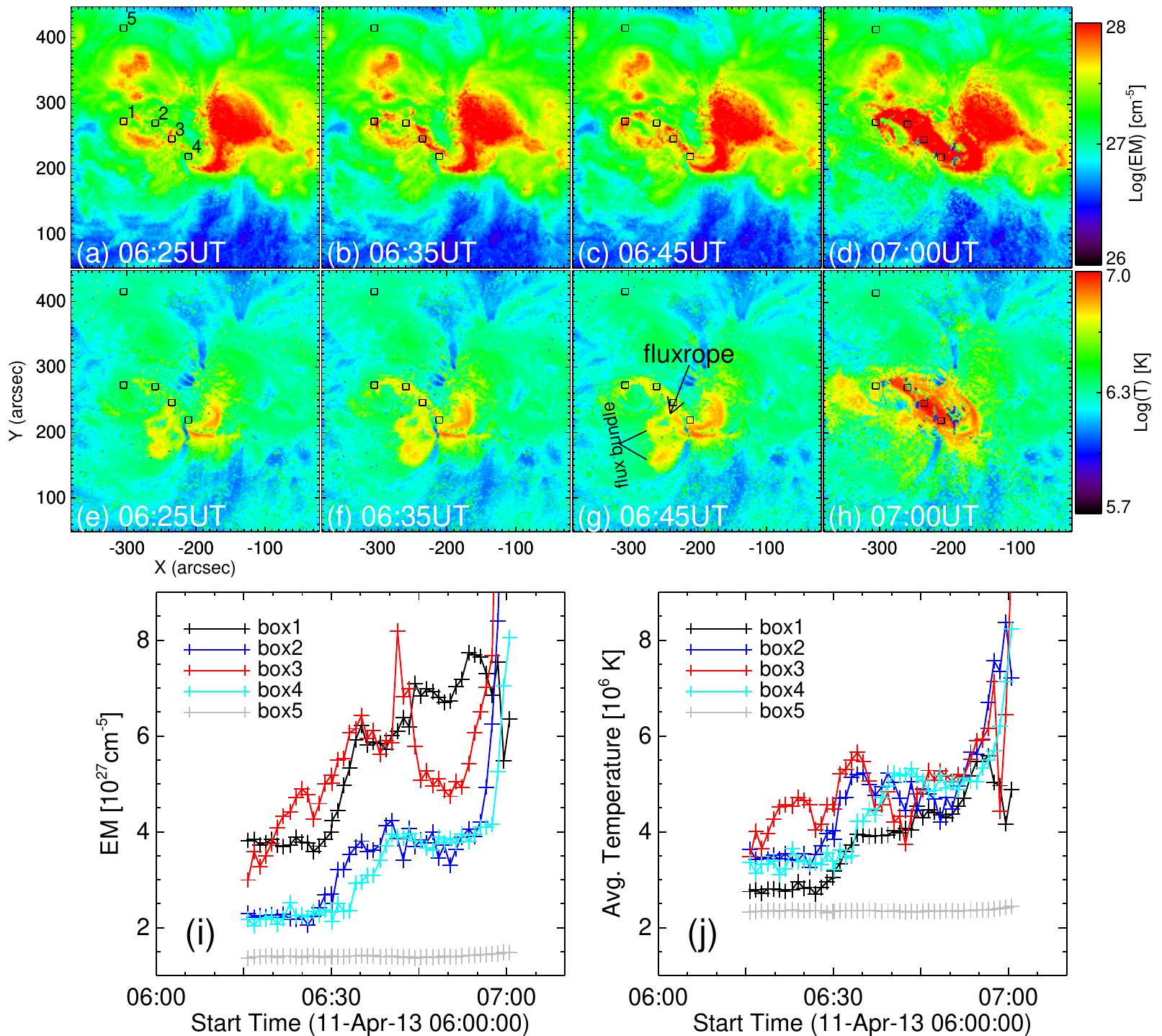}
\caption{Distribution of emission measure (panels (a)-(d)) and temperature (panels (e)-(h)) at different epochs of pre-flare (before the main flare phase) in the AR. For the study of time evolution of these parameters, we considered four squared boxes (1-4) along the FR and another remote box-5 chosen away from the FR. The time evolution of total EM, and temperature $\bar{T}$ averaged in boxes 1-5 are plotted in panels (i) and (j), respectively. A major difference in these profiles from remote box and the boxes along the FR confirms the enhanced emission and temperature due to FR activity. To derive total EM and $\bar{T}$, the DEM is integrated over the temperature range of \textit{LogT}=5.7 to \textit{LogT}=7.1.}\label{Fig_EM_TEMP}
\end{figure*}

Therefore, there exist two phases of slow rise and fast acceleration phases in the kinematics of FR as many previous reports \citep{zhangj2001,zhangj2004}. Note that it took about 30 minutes (06:15-06:45\,UT) for the initiation of FR to trigger the eruption of Earth directed CME. Importantly, the onset time of fast acceleration (06:45\,UT) of the overlying flux bundle is 10 minutes prior to the onset time of the flare. Interestingly, in an eruption event of moderate CME and M1 flare, \citet{Vemareddy2011} found the entire span including slow and fast rise motion phases of FR as 15 minutes which is very small as noticed in this case as 40 minutes. This gives us the impression that strong eruption events have longer activation periods compared to that of weak events, which needs further assertion from a study of many such cases.

\subsection{Thermal Properties}
\label{TherProp}
Thermal structure and evolution of the sigmoid and the presumed FR are explored from differential emission measure (DEM) analysis using six EUV channels from AIA. We used \texttt{xrt\_dem\_iterative2.pro} in SolarSoftWare with modifications that is designed for Hinode/XRT \citep{golub2004,weber2004} data. The code implements forward fitting method to construct DEM at each pixel given its flux and temperature response function in each passband. The validity of this code was extensively tested on AIA data in a recent study of CME thermal components by \citet{chengx2012}. They found capability of the code to reproduce theoretically constructed AR DEM with good accuracy across the entire AR. Following the same methodology, we constructed DEM maps of AR throughout the evolution and evaluated DEM weighted average temperature ($\bar{T}$) and total emission measure (EM) defined as
\begin{equation}
\bar{T}=\frac{\int{DEM(T)\,T\,dT}}{\int{DEM(T)\,dT}};\hspace{.2 in} EM=\int{DEM(T)\,dT}
\end{equation}
Note that the integrations are evaluated over the temperature range $5.7<LogT<7.1$. Given EM, the density at any location can be estimated as $n=\sqrt{EM/l}$ where l is depth (or width) of the pixel assuming the filling factor to be 1.

In Figure~\ref{Fig_EM_TEMP}(a-f), we plot EM and $\bar{T}$ maps depicting the distribution of extended emission measure and average temperature of FR in the AR.  The distribution of EM in quiet regions (of the orders of 26) is well dominated by that in the regions of FR; it reaches to 28 orders ($cm^{-5}$) along the FR. While the emission increases with the onset of the FR rise from 06:30\,UT, this value also enhances towards maximum values of order 29, however we scaled the images for the visibility of small values at better contrast.

Similarly, the derived temperature behaves according to EM. From the figure, (panels~\ref{Fig_EM_TEMP}(e-f), green, yellow, red), we can notice that higher temperature of the FR and the overlying flux bundle as well. As pointed earlier, owing to its appearance in hotter passbands in AIA, the signatures of pre-existing FR gradually becoming apparent by its distinguished higher temperature distribution from the background plasma thermal conditions. The pixels in the FR have the temperature $\sim$2.0\,MK in the initial phases and reaches to 10\,MK as the evolution tending towards onset of eruption and flare with ribbons of bright emission. In total, the FR features exhibit unique structure of density, temperature, and emission that are quite different from background thermal properties.

As the background emission prevents to estimate true values of EM and $\bar{T}$ at each FR pixel, we are more interested in their temporal variation relative to background values, rather than exact values with possible errors. The averaged temperature and EM are evaluated in small regions ($12\times12$\,arcsec$^2$) represented by square boxes (1-4, in panel~\ref{Fig_EM_TEMP}(a)) along the FR (Figure~\ref{Fig_EM_TEMP}(i-j)). Since, we have not subtracted the background emission in the FR, another identical box 5 is selected at quiet region as a representative of background flux, where it is unaffected by extended emission due to FR activity. Obviously, the time evolution of these parameters from box 5 almost remains stable and it is reasonable to regard those from boxes 1-4 as being truly related to external heating mechanism associated within/underneath the FR as explained earlier. Starting from 06:15\,UT, boxes 1, 2, 3 show earlier pickup of these parameters before 06:35\,UT, whereas box 4 shows late phase emission, probably because it is away from rising part of FR and became dominated by the emission in near surface flare region. Interestingly, these time profiles have correspondence with that of rising motion of FR. Slow rise phase till 06:30\,UT has the trend of slow increase of the parameters. After that they decreased because the core of the FR reached up falling away from the chosen boxes in the projected view of observations. However, the fast rise motion commences the flare emission at these boxes resulting in steep increase trend with the abrupt rise of temperature and EM.

For the density estimates, the line-of-sight depth is a factor taken from observations of FR. Usually, it is considered as cross-section size (cut across the FR, or perpendicular to the axis of the FR) of FR. As the FR rises, it expands with increasing cross section size and is a variable with time. At initial phases, the cross section size is found to be about $\sim$8\,Mm and expands to $\sim$23\,Mm just before the onset of the flare. With this, the estimated density from the EM in these boxes along the FR is $\sim$$5.3\times10^9$ cm$^{-3}$ in the initial phase and remains to be at similar value $5.8\times10^9$ cm$^{-3}$ in later times. Although these values are consistent with previously found cases \citep{chengx2012} to the nearest of an order, the later value might be overestimated because the increased emission due to reconnection yields high EM values, which give larger density estimates when compared to the real situation in expanding FR without reconnection below it (i.e., flare contamination to the FR).

In summary, the evolution of these parameters due to rising motion of FR exhibits net increased variation, i.e., EM in the range $\sim(2.3-9.0)\times10^{28}$\,cm$^{-5}$, density in the range $\sim(5.3-5.8)\times10^{9}\,cm^{-3}$, and temperature by $\sim(2.6-9.0)$\,MK, in accordance with predicted FR models of CME which emphasize thinning current sheet underneath the FR in which reconnection heats the surrounding plasma.

\begin{figure*}[!htpb]
\centering
\includegraphics[width=.97\textwidth]{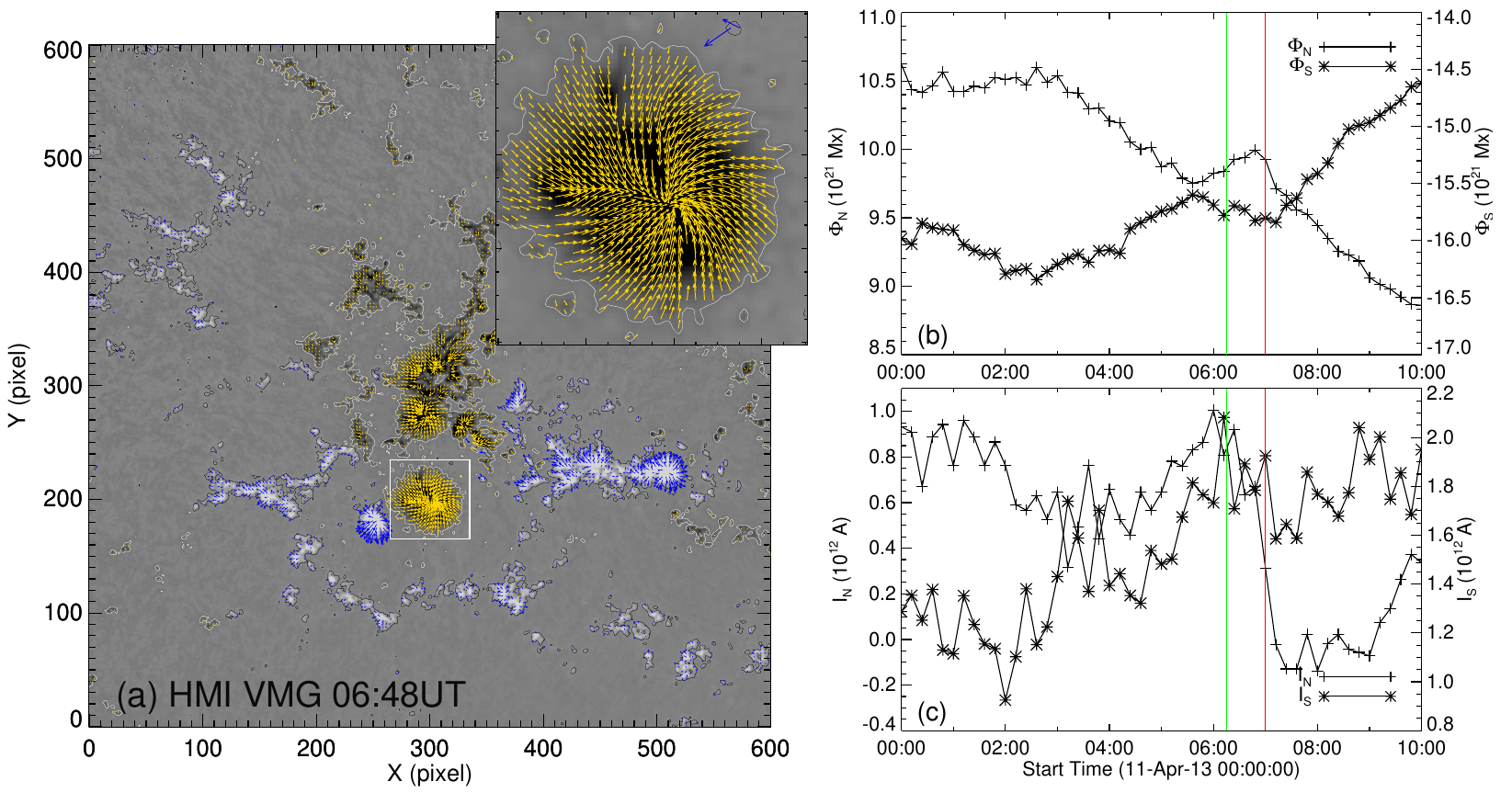}
\caption{Magnetic field and its derived quantities in AR 11719. a) HMI vector magnetogram plotted with horizontal vectors (yellow) on LOS map. For a close view, the sunspot fields are shown in inset. The arrow represents direction and length is proportional to magnitude of horizontal magnetic field.  (b-c) net line of sight flux ($\Phi$), net vertical current (I) obtained by summing over north and south polarities in the entire AR field-of-view as panel (a). Vertical green, red lines refer to the start time of FR rise, and M6.5 flare.}\label{Fig_VMgram_Bf}
\end{figure*}

\begin{figure*}[!htpb]
\centering
\includegraphics[width=.65\textwidth]{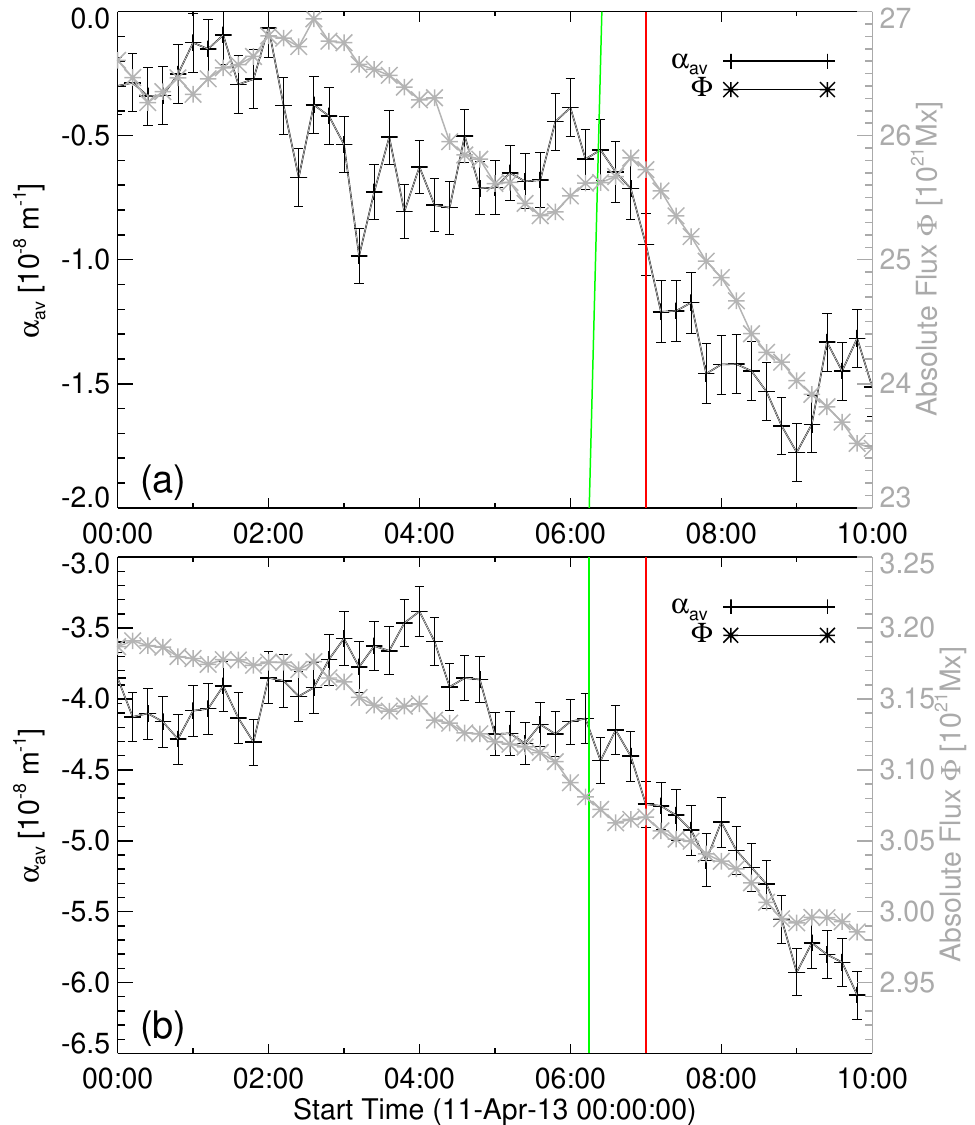}
\caption{Time evolution of $\alpha_{av}$, total absolute flux estimated over a) entire active regions, b) the sunspot region as shown in the inset of panel~\ref{Fig_VMgram_Bf}(a). Vertical green, red lines refer to the start time of FR rise, and M6.5 flare.}\label{twist_para}
\end{figure*}

\subsection{Magnetic Properties}
\label{MagProp}
In this section we have studied magnetic field evolution at the photosphere for their role in FR eruption.

There are weak horizontal field distributed in diffused fluxes as observed in vector magnetograms of the AR (Figure~\ref{Fig_VMgram_Bf}(a)). However, the directions of strong horizontal fields in the sunspot S1 (inset plot) show slightly twisted behavior of sunspot as a whole. As the FR and filament originates from this sunspot, the orientation of the horizontal field vectors is in a manner that the anchored field lines twisted left handed, in agreement with the chirality of FR. In the gradual phases of evolution of these AR magnetic fields, this sunspot anchored FR is driven towards upward rise motion and finally to unstable state. Therefore, it is very necessary to identify the magnetic field evolution and its twisting corresponding to such a transformation of magnetic structure in the AR. Especially, the sunspot fields are very important in that respect.

\subsubsection{Evolution of magnetic flux and net vertical Current}

With the horizontal field components, we can only compute vertical current density $J_z=\frac{{{(\nabla \times \mathbf{B})}_{z}}}{{{\mu }_{0}}}$, where ${{\mu }_{0}}=4\pi \times {{10}^{-7}}Henry\,\,{{m}^{-1}}$. The net line-of-sight flux ($\Phi=\sum\limits_{N}{{{\text{B}}_{\text{z}}}}$) and net vertical current ($\text{I}=\sum\limits_{N}{{{\text{J}}_{\text{z}}}}$) are evaluated over north (N) and south (S) polarities of AR separately, and plotted with time in Figure~\ref{Fig_VMgram_Bf}(b-c). For the constraints of noise, we considered pixels possessing values $|B_z|>50$\,G and $|B_t|>150$\,G. Both the fluxes showed slight increase in the first two hours and then showed continuous decreasing trend till 10:00\,UT on April 11. Within 8 hours duration, the net decrease in the fluxes is $\sim 1.3\times10^{21}$\,Mx which is far higher than that due to fluctuations, therefore it can be considered to have influence on surrounding fields and the field within the FR that is embedded in AR flux system. Moreover, the net vertical current from south polarity $I_S$ exhibited drastic increase from 1 hour prior to onset time of FR rise. Even net vertical current $I_N$ over north showed increased behavior in the same time. Due to fluctuations in the profile, we can take the average level before and after the onset time of FR rise to quantify the net change in the profile. The net change of $I_S$ ($I_N$) before and after the FR onset time estimated as $0.9\times10^{12}$\,A ($0.6\times10^{12}$\,A) which is more than the fluctuations of maximum amplitude $0.5\times10^{12}$\,A. This increase of net current is generally considered to be precursor of the occurrence of eruptive events like flare and CME \citep{vemareddy2012b, ravindra2011, schrijver2009}. Further, we can notice that the north polarity flux (south polarity flux) is having dominant negative net current (positive net current) which implies negative signed or left handed chirality in the AR flux system \citep{wangj2004}, that is consistent with the magnetic channel of inverse-S FR.

\subsubsection{Evolution of magnetic twist}

The helicity or twist of the AR can be estimated by the proxy \textit{average alpha} (${{\alpha}_{av}}$). By computing the parameter $\alpha$ from the force-free assumption of photospheric fields, i.e., $\nabla \times \mathbf{B}=\alpha \mathbf{B}$, we estimated ${{\alpha}_{av}}$ \citep{pevtsov1994,hagino2004} given by

\begin{equation}
{{\alpha}_{av}}=\frac{\sum{{{J}_{z}}(x,y) \text{sign}[{{B}_{z}}(x,y)]}}{\sum{|{{B}_{z}}|}}
\label{Eq_Alp_av}
\end{equation}

The error in ${{\alpha}_{av}}$ is deduced from the least squared regression in the plot of $B_z$ and $J_z$ and is given by
\begin{equation*}
\delta \alpha _{av}^{2}=\frac{{{\sum{\left[ {{J}_{z}}(x,y)-{{\alpha }_{av}}{{B}_{z}}(x,y) \right]}}^{2}}/|{{B}_{z}}(x,y)|}{(N-1)\sum{|{{B}_{z}}(x,y)}|}
\end{equation*}
where N is the number of pixels with $|B_t|>150$\,G.


The time evolution of ${{\alpha}_{av}}$ over the entire AR and the sunspot region (inset in panel~\ref{Fig_VMgram_Bf}(a)) is plotted in Figure~\ref{twist_para}(a-b). The sign of ${{\alpha}_{av}}$ indicates negative helicity or chirality that is consistent with the aforementioned interpretations from morphological study and net current as well. Predominant increasing trend of ${{\alpha}_{av}}$ from about 06:00\,UT onwards indicates increasing twist of the AR flux system. However, the similar increasing trend over sunspot starts early from 04:00\,UT onwards. Note that the magnitudes of ${{\alpha}_{av}}$ in sunspot is about four times to that in entire AR as a fact of averaging effect in large area of unevenly distributed horizontal fields.

\subsubsection{Possible role of kink instability in triggering the eruption}
The question for us to ask is whether this constant increase of twist has any effect in triggering the eruption? According to kink instability criteria, if the FR twist increases beyond a certain threshold, it is prone to unstable kinking nature and writhing. The exact amount of twist required for a kink instability depends on several factors, including the loop geometry and neighboring/overlying fields, but is generally agreed to be at least one full wind \citep{hood1979,baty2001,torok2004,fany2003,fany2004}.

To our knowledge, \citet{leamon2003} were the first to estimate the total twist T of the coronal magnetic loop assuming as a semicircle of length l with its footpoint separation distance d as given by
\begin{equation}
T=lq=\frac{\pi d}{2}\frac{{{\alpha}_{best}}}{2}\label{Eq_wind}
\end{equation}
\noindent where ${{\alpha}_{best}}$ is force-free twist parameter similar to ${{\alpha}_{av}}$. Their results of the study inferred small twist of CME-producing ARs and thus inconsistent with kink instability to be a cause of solar eruptions. However, introducing a different parameter ${{\alpha }_{peak}}$ obtained at localized portion of AR instead of ${{\alpha }_{best}}$ over the entire AR, \citet{leka2005} argued availability of sufficient twist in AR FRs that is greater than $2\pi$.

The distance between two legs of FR that lie at S1 and N2 is 66\,Mm. Since the ${{\alpha}_{av}}$ [for the isolated sunspot region] increases from -3.38 $(\pm0.35)\times10^{-8}$\,m$^{-1}$ at 04:00 UT to -4.16 ($\pm0.32)\times10^{-8}$ m$^{-1}$ at 06:00\,UT, the estimated twist of FR that originates from sunspot lies in the range $0.28<T/2\pi<0.34$ turns with an error bar of 0.02 turns which does not satisfies the threshold criteria for kink instability. Note that, in the expression of equation~\ref{Eq_wind}, the winding rate (q) is assumed to be half of the force-free twist parameter in the case for the straight cylindrical flux tube. In general the relation between winding rate along the axis of the tube and the twist parameter is not well known.

These conditions are very differed by a factor of three when considered ${{\alpha }_{av}}$ from entire AR, nevertheless it exhibits increasing trend from before the onset of the eruption. The very basic puzzle strike to us is from where the observed twist of flux bundle and the FR originated? For the FR of semi-circle geometry of length 104 Mm, the minimum required twist from the source active region is $1.21\times10^{-7}$m$^{-1}$. At this value the FR will have one turn ($2\pi$) of twist. Although the distributed $\alpha$ ranges between $\pm10^{-6}$\,m$^{-1}$, the averaging effect brings down the value by about an order depending on the extent of area considered and presence of strong shear field distribution.

\begin{figure}[!htpb]
\centering
\includegraphics[width=.49\textwidth]{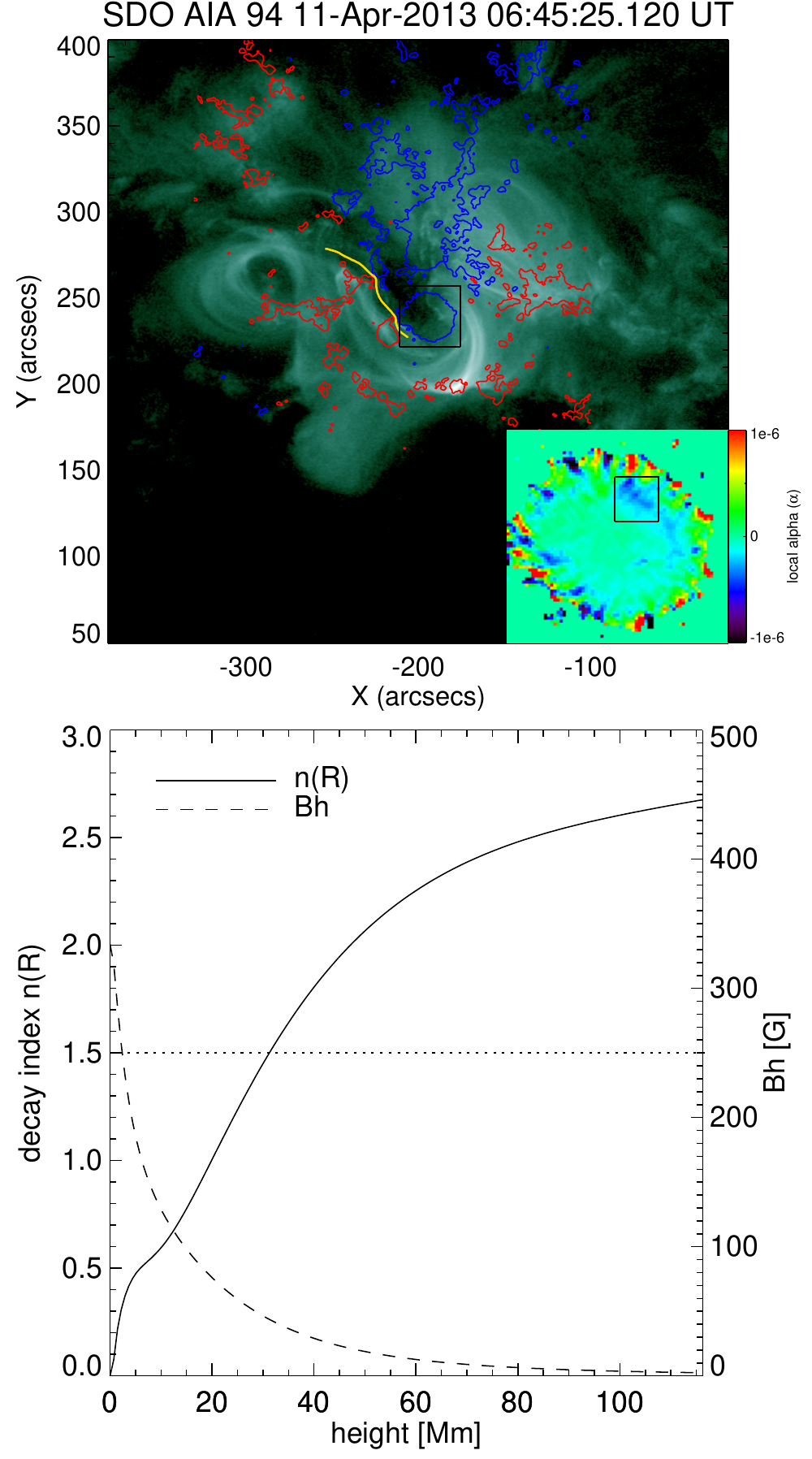}
\caption{\textit{Top panel} Snapshot of corona in AIA 94\,\AA~passband showing FR structure with plasma loops. Contours of line-of-sight magnetic 	flux at $\pm150$\,G are overlaid in red/blue color to indicate link between photospheric magnetic flux and coronal FR structure. The distribution of force-free parameter $\alpha$ ($=J_z/B_z$) of the main sunspot region (marked by black square box) is depicted in the inset plot. The region outlined by square box in the inset plot indicate negative signed alpha having critical average value for kink instability criteria. The gold color curve is PIL. \textit{Bottom panel} The decay index of $B_h$ as function of height. Also shown is $B_h$ with y-axis scale on right. The dotted horizontal line refers to critical decay index value.}\label{Fig_Kink}
\end{figure}

We have scrutinized the association of photospheric magnetic flux and coronal FR structure by plotting line-of-sight flux contours on 94~\AA~passband image as depicted in Figure~\ref{Fig_Kink}. As described earlier, the FR leg launches from sunspot S1 and lands at positive flux region N2. Note that one of the footpoint of the twisted flux bundle rooted from positive polarity near N2 and there is no negative polarity flux (contours) near the sunspot, the other leg must be anchored in the periphery of the sunspot S1. This different spatial location of the flux bundle and the FR within the sunspot delineates to have unequal amounts of twist and so different flux systems. The west part of the sunspot S1 is more associated to FR leg orientation and is believed to have significant contribution to the twist of the FR. It is also supported by the vector field plot (Figure~\ref{Fig_VMgram_Bf}a) which shows asymmetric shear distribution throughout the sunspot S1. These vectors follow the FR orientation at that west part of the sunspot. The alpha distribution is monitored in this sunspot region (inset plot of Figure~\ref{Fig_Kink}). The peripheral parts of sunspot contain small locations with $\alpha$ more than $10^{-6}$ m$^{-1}$ with both signs. Evidently, we can identify the patch of negative $\alpha$ which is having values of order -7. The average value of $\alpha$ in a small region (shown in square box) meets the threshold criteria for helical kink instability for this FR whose leg originates at this portion of the sunspot. The value of $\alpha_{\rm av}$ is slightly below the critical value ($1.21\times10^{-7}$\,m$^{-1}$) at 04:00\,UT and crosses at the onset time of the eruption. The same exercise cannot be carried out for the flux bundle as we cannot pinpoint the footpoint location of it within the sunspot. Thus, the monotonous increase of the twist of the AR flux system to a exceeding value may support the possibility of observed kink evolution of the overlying flux bundle, despite the involved projection effects (Section~\ref{MorphProp}) and primarily the rise motion of the flux rope.

According to Torok-Kliem theory \citep{torok2004}, the kink instability can only cause the rise motion of the FR at the kinking portion. It cannot produce the eruption itself. The further eruption depends on three possible mechanisms: (1) the triggering of the torus instability (the fast declining of the overlying field reaches a critical value at the new height), (2) the triggering of the magnetic reconnection in the current sheet induced by the kink-rise motion, (3) the triggering of magnetic reconnection with the neighboring loops. 

\subsubsection{Background magnetic field and the torus instability}

As required by the torus instability which is the supposed driving mechanism (case 1), the background magnetic field gradient overlying the FR must exceed certain critical value \citep{torok2005,olmedo2010,chengx2011}. For this, we have calculated the background field decay index given by
\begin{equation} 
n(R)=-\frac{\text{R}}{\text{B}}\frac{\partial \text{B}}{\partial \text{R}}
\end{equation}
where R is the geometrical height from the boundary point and B is the field strength. Since the field component parallel to the rising direction of the FR (vertical field component, $B_z$) does not contribute to inward confining force, the decay index is generally calculated for the perpendicular component (horizontal field component, $B_h$). We computed the background field in the entire volume of FR structure by potential field approximation using the observed vertical component of magnetic field at the photosphere.  The decay index of $B_h$ and $B_h$ as a function of height from a point on the polarity inversion line is plotted in Figure~\ref{Fig_Kink} (top panel). Starting from 0 at the photosphere, the decay index reaches to 2.7 at a height of 117\,Mm. From FR models, the critical value of the decay index generally found to be in the range 0 to 2 and varies depending on the fraction of the torus above the surface. \citet{torok2005} proposed a constant value of 1.5 as a critical decay index. The extrapolated field reached this critical value at a height of 31\,Mm. At about 10\,Mm the curve exhibits a gradual steepness which is indicative of ``bump'' as found by \citet{chengx2011} interpreting as a distinct shape for eruptive and non-eruptive flare events. The $B_h$ decreases faster in the low corona as noticed (about 10\,Mm), the appearance of ``bump'' in the eruptive flare cases may indicate easiness of the FR to experience torus instability. From our result, the sigmoid probably lying in the lower corona (about 10\,Mm) have to rise slowly overcoming the so-called ``strapping effect'' of horizontal field till 31\,Mm of height, from where this effect weakens exponentially beyond a limiting value falling into a situation like torus instability.

Therefore, these conditions of magnetic fields suggest that the AR flux system had ample twist for the kink instability. It initiates rise motion of over-lying flux bundle, and allows the under-lying sigmoid (flux rope) to rise to the critical point of torus instability. In addition, we also found evidence for the availability of critical twist in the FR itself, which might also have helped in overcoming the strapping effect of the horizontal field within the low corona.

\section{Summary and Discussion}
\label{Summ}
We have investigated an eruption event of Earth directed CME involving sigmoidal FR. A morphological study for the existence of the FR and features of magnetic structure based on EUV observations, DEM analysis for evolutionary investigation of density, temperature structure of FR, and a study of AR magnetic fields for their role in FR eruption are presented. The results of these studies are inferred based on current understanding of CME eruption models and are consistent with most observations. 

The eruption event occurred on 11 April, 2013 in the solar AR 11719 at 06:45\,UT. The eruption source region in the active region contains multiple twisted magnetic flux bundles stacking on each other. The main part is the inverse S-shaped sigmoid seen in AIA hot passbands, presumably to be the magnetic FR of this complex system. Overlying the sigmoid, there exists a twisted flux bundle, which also appears in hot passbands. A filament, appearing in cool temperatures, lies below the sigmoid.

From 06:06\,UT onwards, the overlying twisted flux bundle exhibits structural evolution that transforms from V- through $\gamma$- to \textgamma-shape. This kind of transformation in the geometrical shape is similar to kink-rise as a signature of magnetic twist in accordance with conservation of helicity. The exceeding twist in the FR of a particular chirality (positive/negative) begins to wound the axis of the FR around itself. Beyond a critical value, the FR (here flux bundle) is subjected to kink instability. By the helicity conservation rule, a part of the twist is converted to writhe of same handedness (left/right) as evidenced here \citep{torok2004}. A kinematic study reveals that rise motion of this flux bundle consists of slow rise and relatively fast acceleration phases which are temporally coincide with that of the main FR. The later fast acceleration phase is followed by eventual eruption of FR and associated two ribbon flare from 06:55\,UT onwards.

The upward rise motion of twisted flux bundle overlying the FR is co-temporal with that of FR. We interpret that the observed $V-\gamma-$\textgamma~transformation of flux bundle is a signature of magnetic twist experiencing helical kink instability of surpassing twist.  And its co-temporal rise motion with FR implies that either it allows the underlying FR to rise and/or the critical twist in the FR makes it to rise (Figure~\ref{Fig_Kink}). All of these inferences imply the role played by kink instability of magnetic twist in the AR flux system and that in turn trigger the eruption as the upward rise motion of the FR. As the kink instability itself cannot produce eruption, the further eruption mechanism is likely driven by onset of the torus instability \citep{torok2004,torok2005}, resulting in the Earth-directed CME.

With the rise motion of FR, bright emission proportionately increases which gives the appearance of sigmoidal flux system in coronal hot imaging channels of AIA. The heating of plasma by persistent reconnection in the current sheet underneath the rising FR is likely the mechanism for the brightness of this sigmoidal system, as predicted by numerical simulations \citep{titov1999, gibson2006}. The DEM analysis of FR unveils the variation of thermal properties that have correspondence with the FR rise motion. Although the background emission restricts us to estimate true values, the calculated thermal properties revealed the FR activity relatively over those of background. The time evolution of the parameters due to rising motion of FR exhibits net increased variation, i.e., EM in the range $\sim$$(2-10)\times10^{28}cm^{-5}$, density in the range $\sim$$(5.3-5.8)\times10^9\,cm^{-3}$, and temperature by $\sim$(2.5-8.0)\,MK.

The net magnetic flux of the AR evaluated at north and south polarities showed decreasing behavior and the net current in them exhibits increasing trend. As the positive (negative) flux is having negative (positive) current, the chirality of AR flux system is negatively (left handed) twisted which is consistent with the chirality of inverse S-sigmoidal FR. Under these field conditions, the time profile of twist parameter $\alpha_{\rm av}$ is increasing from about 2 hours prior to onset of FR rise which states that AR flux system is being twisted during the decay or cancellation of the distributed fluxes and supports the shape transformation of the flux bundle due to magnetic twist.

Especially, the time profiles of flux and $\alpha_{\rm av}$ deduced from the sunspot infers a similar conclusion. As the FR anchors from the sunspot, we estimated the twist number by assuming the FR as a half torus \citep{leka2005}. The obtained values indicate that the twist in the FR is strong enough to meet the threshold criterion (more than one turn) for the helical kink instability. The kink-rise motion of the flux bundle and the critical twist of the FR together played crucial role in the slow rise motion of the FR to a critical point of the torus instability.

The major point to be discussed is how the FR structure formed and how it acquired twist to a critical point. The AR fields in which the sigmoid is embedded are decreasing in time in both polarity fluxes. This decrease of fluxes can be regarded as flux cancellation in which small scale opposite polarity magnetic fragments disappear. There have been evidences \citep{yurchyshyn2001,bellotrubio2005} that the reconnection location being associated with the photospheric flux cancellation. This reconnection and cancellation of magnetic flux is of specific interest, based on which \citet{balleggoijen1989} proposed a mechanism for the formation of helical field lines from the sheared arcade along the polarity inversion line and also build-up of FR. Indeed, the photospheric flux cancellation is evidently reported to be associated with FR formation and its eruption \citep{green2009, tripathi2009, green2011}.

In our case, the FR structure is seen to be already formed by the beginning of April 11. This well augmented FR in the AR flux system of further decreasing fluxes would likely indicate its build-up by increasing twist. Note that the torsion parameter from the sunspot predominantly increases well before the eruption onset. Therefore, this analysis of magnetic fields of source AR suggests that the canceling fluxes are prime factors to the monotonous twisting of the FR system reaching to a state at which the twist is sufficient to trigger kink instability to initiate rise motion. This subsequently triggers torus instability ultimately to eruption of Earth directed CME and reconnection in the current sheet underneath it leads M6.5 flare.

\acknowledgements We thank an anonymous referee for the constructive comments and suggestions. The data have been used here courtesy of NASA/SDO and HMI, AIA science teams. We thank the HMI science team for making available processed vector magnetogram to the solar community. The author is indebted to Dr. Xin Cheng for his programmable help on DEM computation. J.Z. is supported by US NSF ATM-0748003 and NSF AGS-1156120.

\bibliographystyle{apj}

\end{document}